\documentclass[10pt, conference, compsocconf]{IEEEtran}
\usepackage[]{amsmath}
\usepackage{amssymb}
\usepackage{xspace}
\usepackage{url}			
\usepackage{booktabs}
\usepackage{wrapfig}
\usepackage{enumerate}
\usepackage{todonotes}	
\usepackage{subfigure}	
\usepackage{lipsum} 		

\usepackage{listings}
\definecolor{lst_green}{rgb}{0,0.6,0}
\definecolor{lst_gray}{rgb}{0.5,0.5,0.5}
\usepackage{ntheorem}

\newcommand{\vbar}{\;|\;}

\usepackage{longtable}
\newcommand{\beginLongtable}{\begin{longtable}{p{5.2cm}p{9.8cm}}}

\newcommand{\Veh}{\mathbb{V}}

\newcommand{\SetN}{\mathbb{N}}

\newcommand{\toSetN}{\rightarrow \SetN}

\newcommand{\SetBool}{\{0,1\}}

\newcommand{\toSetBool}{\rightarrow \SetBool}

\newcommand{\toSet[1]}{\rightarrow \{{#1}\}}

\theoremstyle{break}

\theoremstyle{plain}

\newtheorem{define}{Definition}

\graphicspath{{./figures/}}

\def\figureWidthSmall{0.4\textwidth}

\ifCLASSINFOpdf
\else
\fi

\begin{document}

%
\title{Deployment Calculation and Analysis for a Fail-Operational Automotive Platform}

\author{
\IEEEauthorblockN{Klaus Becker, Bernhard Sch{\"a}tz, Christian Buckl}
\IEEEauthorblockA{fortiss GmbH\\
Guerickestr. 25, 80805 Munich, Germany\\
Email: \{becker, schaetz, buckl\}@fortiss.org}
\and
\IEEEauthorblockN{Michael Armbruster}
\IEEEauthorblockA{Siemens AG, Corporate Technology\\
Otto-Hahn-Ring 6, 81739 Munich, Germany\\
Email: michael.armbruster@siemens.com}
}



\maketitle


\begin{abstract}

In domains like automotive, safety-critical features are increasingly realized by software. Some features might even require fail-operational behavior, so that they must be provided even in the presence of random hardware failures. A new fault-tolerant SW/HW architecture for electric vehicles provides inherent safety capabilities that enable fail-operational features.

In this paper, we introduce a formal model of this architecture and an approach to calculate valid deployments of mixed-critical software-components to the execution nodes, while ensuring fail-operational behavior of certain components. Calculated redeployments cover the cases in which faulty execution nodes have to be isolated. This allows to formally analyze which set of features can be provided under decreasing available execution resources.

\end{abstract}



%
\IEEEpeerreviewmaketitle

\section{Introduction and Motivation}
\label{sec:intro}

Embedded systems are often operated in safety-critical environments, 
in which unhandled faults could cause harmful system failures.
Hence, safety-critical systems have to react on faults properly.
Many current safety-critical systems for mass-markets, like vehicles, handle faults by invalidating faulty data and avoiding harm by going into a fail-safe state.
However, this may cause the loss of provided features. 
This is not acceptable for features that require fail-operational behavior.

To increase their dependability, systems must be able to resume affected features without any service interruption.
If system resources get lost due to hardware failures, runtime-reconfiguration can be applied to efficiently use the remaining resources.
As the remaining resources may become insufficient to provide the full set of features, the explicit deactivation of some features would allow to keep alive the subset of features with the highest demand with respect to safety, availability and reliability. We use these terms as defined in \cite{avizienis2004basic}.

However, in current automotive E/E architectures, reconfiguration is substantially restricted by Electronic Control Units (ECUs) tailored to their provided features, and inflexible communication buses. This heterogeneity prevents a system-wide mechanism to increase the feature reliability by resuming relevant software on other ECUs after hardware failures. Additionally, the integration of new features becomes more and more complicated due to increasing feature interactions.
These and more challenges are for instance discussed in \cite{chakraborty2012embedded}. It is stated that a substantial revision of the vehicles HW/SW architecture can reduce its complexity to an adequate level.

We propose a new centralized HW/SW platform for vehicles, capable to overcome the mentioned shortcomings. 
The platform provides inherent safety properties and supports fail-operational features without requiring mechanical fallbacks. 
To avoid harm, faulty hardware is isolated from the remaining system. Affected software is resumed on intact hardware. 
Extensions to the feature-set after sale are supported in a Plug-and-Play manner.

In this paper, we address the calculation and analysis of the deployment of software components to the execution nodes inside the proposed architecture. To provide fail-operational features, software components are deployed redundantly. 
However, with a rising number of software and hardware units, this configuration becomes more and more complex and hard to manage manually. We therefore provide an automated configuration support for deployment decisions, ranging from a semi-automated to a fully-automated approach. 
Our approach is based on a formal system model and a set of formal constraints that describe the validity of deployments with respect to the safety-concept. Model and constraints characterize an arithmetic problem that can be solved by SMT-solvers.

The main contribution is an approach to calculate and analyze different reconfigurations of the deployment to become active after execution nodes become isolated. 
The set of active software components -- and thus also the set of provided features -- is automatically reduced when the remaining system resources become insufficient to provide the initial set of components. 
Components are deactivated based on their priorities, which can either be assigned manually or derived automatically.
Our approach allows to formally analyze at design-time if the desired system and feature properties can be fulfilled, like which set of features can still be provided after one or multiple isolations. 
Analyzing the deactivations of single features allows to analyze the entire system degradation. 

In section \ref{sec:raceFoundations} we present the basic concepts of the proposed platform.
Section \ref{sec:deployment} shows the main contribution of this paper, which is a formal model and a constraint-based approach to calculate valid deployments and to analyze which features can be provided after isolations of execution nodes. 
The applicability is shown by a little example from the automotive domain.
Related work is discussed in section \ref{sec:rw} and the conclusion and future work is given in section \ref{sec:conclusion}.

\section{System Architecture and Safety Concept}
\label{sec:raceFoundations}

\subsection{System Architecture}
\label{subsec:sysArch}

In this paper, we discuss the deployment calculation for a scalable and uniform platform that was developed with the aim to reduce the complexity of automotive HW/SW architectures.
The main platform characteristics are:

\begin{itemize}
\item A simplified hardware \& network structure with a scalable set of execution nodes.
\item A homogeneous communication system for the highly available transfer of critical real-time data based on industrial standards (e.g., Ethernet).
\item Integrated mechatronics components (smart actuators and sensors), such as wheel hub motors with integrated steering, braking and damping.
\item A runtime environment (RTE) that can execute both highly available, safety-critical software as well as non-safety-critical functions side-by-side. 
\item Plug-and-play capability to update or extend the vehicles safety-critical features
by retrofit new software and modern sensors/actuators after purchase. 
\end{itemize}
 
\begin{figure}[h!]
\centering
\includegraphics[width = \figureWidthSmall]{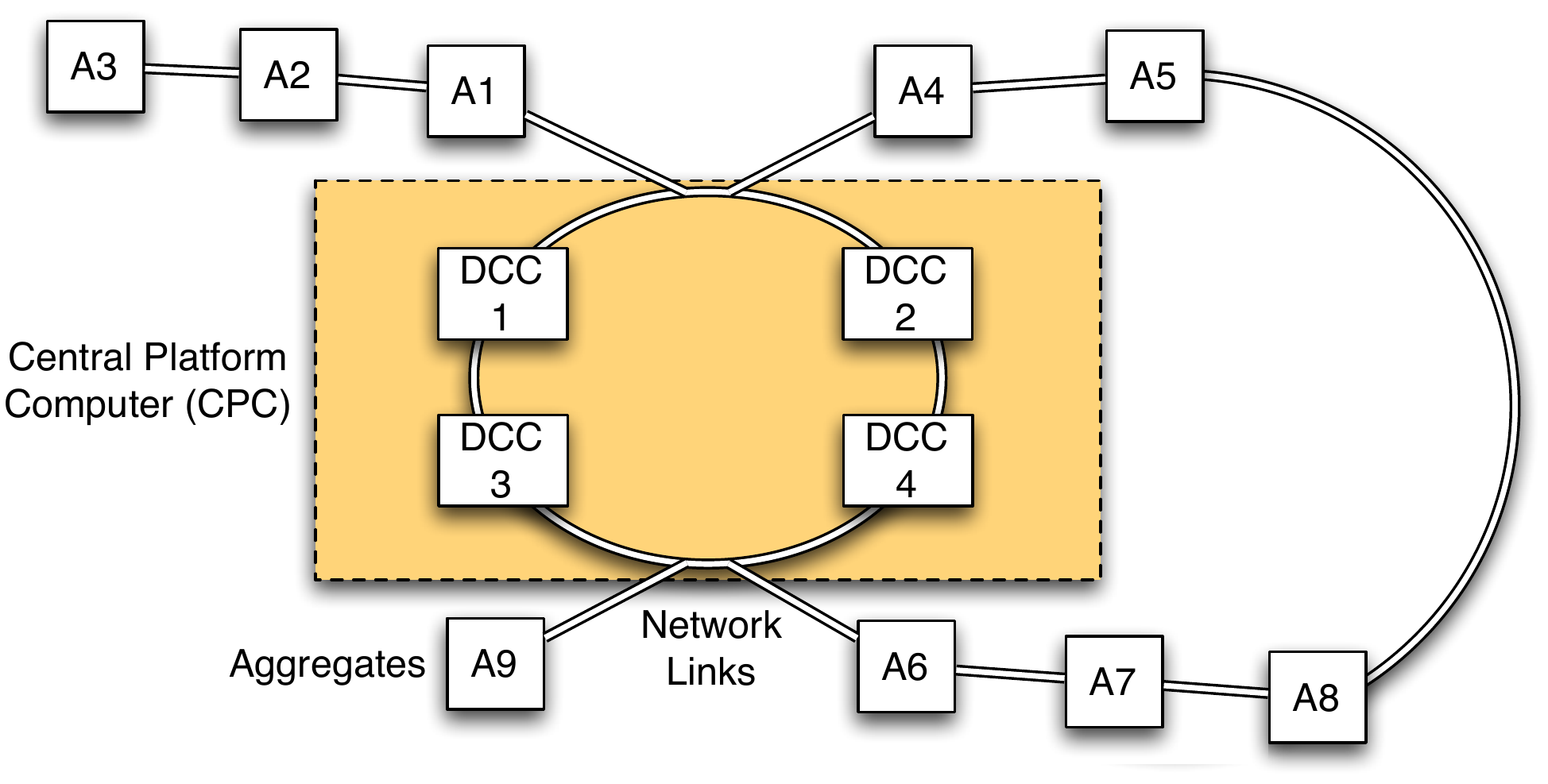}
\caption{Example instance of the proposed hardware architecture}
\label{fig:RACE_Platform}
\end{figure}

The vehicles hardware architecture is composed by a scalable set of central execution nodes (also called \emph{Duplex Control Computers} (DCCs)) and a set of peripheral execution nodes providing the physical sensing and actuating (also called Smart-Aggregates).
The DCCs are connected to each other and to the Smart-Aggregates by redundant switched Ethernet-Links. 
The DCCs assemble the \emph{Central Platform Computer} (CPC). 
We assume homogeneous DCCs for flexibility in the deployment.
Fig. \ref{fig:RACE_Platform} shows an example system architecture. 

The proposed system has two different power supplies \emph{red} and \emph{blue}. 
Each execution node is supplied by either the red or the blue one. 
Hence, if one power-supply gets lost, only a subset of the execution nodes gets lost and the residual nodes can continue the operation. 
These basic properties have also been described in \cite{sommer2013race}.
As scheduling policy, we follow the concept of logical execution times \cite{henzinger2001giotto}, meaning that the software components are executed within \emph{cycles}. 
Each execution node provides a certain budget of time per cycle that can be used to execute application software components. 
In this paper, we assume a simplified model in which all software components are scheduled with the same rate in each cycle.

\subsection{Fault-Model}
\label{subsec:faultModel}

According to ISO 26262 \cite{iso26262_1}, we use the following terms.
1) \emph{Fault}: abnormal condition that can cause an element or an item to fail, 
2) \emph{Error}: discrepancy between a computed, observed or measured value or condition, and the true, specified, or theoretically correct value or condition,
3) \emph{Failure}: termination of the ability of an element to perform a function as required,
4) \emph{Fault-Model}: representation of failure modes resulting from faults,
and
5) \emph{Random Hardware Failure}: failure that can occur unpredictably during the lifetime of a HW element and that follows a probability distribution.

In this paper, we consider random hardware failures that lead to isolations of execution nodes.
According to e.g. \cite{kopetz2003fault}, we further consider Fault-Containment Regions (FCR) as the set of subsystems that share one or more common resources and that can be affected by a single fault.

We consider the vehicle as a set of FCRs. All FCRs have precisely specified linking interfaces in the domains of time and value including a link-specific fault-model. 
We describe the link-specific fault-models per FCR as far as it is helpful to understand the deployment model, presented in section \ref{sec:deployment}.
The relevant FCRs are 
1) execution nodes including its communication-links to data sinks,
2) application software components including its communication-links to data sinks,
and 3) power supplies.

We define for each FCR a fault-model with the states \emph{correct} $z_c$, \emph{faulty} $z_{ft}$, \emph{faulty but passivated} $z_{fp}$ and \emph{faulty out of control} $z_{ooc}$. A random hardware failure with a failure-rate $\lambda$ leads to the transition from $z_c$ to $z_{ft}$. The failure detection and passivation mechanisms within the faulty FCR but also within the receiving FCR lead to a transition from $z_{ft}$ to $z_{fp}$. A passivated FCR does no longer harm the system operation. Anyhow, the functionality of the passivated FCR will no more be available. Only in case that the failure cannot be detected neither by the faulty FCR itself nor by the receiving FCR, we assume that this faulty FCR behaves fully out of control and thus, a correct system-operation can no longer be ensured. 
Out of control $z_{ooc}$ means that the faulty FCR can neither be passivated nor controlled. 
With regard to ISO26262 this is equivalent to the fact that any safety-goal can no more be reached.

In this paper, we assume a state-transition time of $0s$ and a sufficient failure detection coverage. 
Sufficient means that the probability of any FCR to be in state $z_{ooc}$ will be acceptably low to meet the quantitative safety-requirements of the ISO26262 \cite{iso26262_1}. 
Only if these assumptions are true, the deployment-considerations shown later in section \ref{sec:deployment} can be applied in a reasonable manner.

\subsection{Safety \& Redundancy Concept}
\label{subsec:redundancyConcepts}

Fault-tolerance is the ability of a system to maintain control objectives, despite the occurrence of a fault \cite{blanke2001concepts}. 
To achieve this, we deploy multiple instances of application software components in a redundant manner to the execution nodes. 
This enables the system to absorb loss of execution nodes and results in features being fail-operational, 
meaning that features can continue operation in the presence of a limited number of random hardware failures.

In the safety concept of the proposed platform, application software components (ASWCs) are grouped to so called \emph{ASWC-Clusters}. 
These clusters get deployed to the execution nodes of the system. 
Those ASWCs belong to the same Cluster that have the same \emph{Automotive Safety Integrity Level} (ASIL) and the same requirements to behave \emph{fail-operational}. 

Each ASWC has multiple safety goals, while each safety goal has an assigned \emph{fault-tolerance time} ($FTT$).
The smallest of these $FTTs$ is the so called $minFTT$ of an ASWC. 
Likewise, the $minFTT$ of an ASWC-Cluster is the smallest $minFTT$ of the ASWCs that are mapped to this cluster.

Each cluster has at least one deployed instance that is active as a so called \emph{master}. 
If the cluster is required to be fail-operational, a second instance is deployed as a hot-standby or cold-standby \emph{slave} (also known as hot/cold spare). 
The decision to create a hot- or a cold standby slave depends on the $minFTT$ of the cluster compared to the \emph{fault-recovery time} (FRT) of the proposed platform. 

We neglect here the time that is required to switch a cold-standby slave to become a master.
With the proposed platform, a maximum switchover time can be verifed. We actually aim on a switchover-time of max 50ms. In this paper we assume the FRT to be a defined constant as it can be shown that a max FRT can be proven. Due to asynchronities and different fault-detection times depending on the faulty FCR, the actual FRT can be less than the maximum value we assume herein.

Depending on the required level of fail-operationality, additional inactive instances of a cluster are deployed, meaning that they are only in memory but not executed. Hence, we differ between \emph{activations} (active deployments, ASWCs are executed) and \emph{allocations} (inactive deployments, ASWCs are not executed, only in memory). An inactive allocation may become active if this is required after isolations.

Different constraints have to be fulfilled by a deployment to be valid.
For instance, if an ASWC-Cluster has a master and a hot-standby slave, master and slave have to be deployed onto two execution nodes with different power-supplies to avoid that both instances get lost simultaneously when a power-supply fails. 
If the execution node of the master gets isolated, the slave becomes the new master and if required, a passive instance becomes the new hot-standby slave. 

\section{Deployment Calculation and Analysis} 
\label{sec:deployment}

We define the system properties and the deployment problem as shown in the following sections.

\subsection{Formal System and Deployment Model} 
\label{subsec:formalModel}

\begin{define}\label{def:Veh}
A Vehicle $\Veh = \langle F, S^A, H^A, \Phi \rangle$ comprises a set of \emph{Functional Features} $F$, an \emph{Application Software Architecture} $S^A$, an \emph{Execution Hardware Architecture} $H^A$ and a \emph{Configuration} $\Phi$.
\end{define}

\begin{define}\label{def:ASWC}
An Application Software Architecture $S^A = \langle S, SC \rangle$ is composed by a set $S = \lbrace s_1 ,..., s_n \rbrace$ of Application Software Components (ASWCs) and a set $SC = \lbrace sc_1, ..., sc_q \rbrace$ of ASWC-Clusters with $sc_i \subseteq S$ while $\forall i, j: sc_i \cap sc_j = \emptyset$ and $\bigcup_{\, i=1}^{\, q} sc_i = S$. 
We describe the mapping of $s \in S$ to $sc \in SC$ with
$\alpha(s) \xrightarrow[]{} \lbrace sc_i \in SC \vbar sc_i$ contains $s \rbrace$
and
$\alpha(sc) \xrightarrow[]{} \lbrace s_i \in S \vbar s_i$ is mapped to $sc \rbrace$.
\end{define}

\begin{define}\label{def:Feature}
The set of functional features $F = \lbrace f_1 ,..., f_m \rbrace$ contains the features of the vehicle that can be recognized by the user.
A feature is realized by one or more ASWCs and the involved Sensors and Actuators, while each ASWC contributes to realize one or more features.
For $s \in S$ and $f \in F$, we define this relationship as
$\chi(s) \xrightarrow[]{} \lbrace f_i \in F \vbar s$ contributes to realize $f_i \rbrace$ 
and
$\chi(f) \xrightarrow[]{} \lbrace s_i \in S \vbar f$ is partly realized by $s_i \rbrace$. 
\end{define}

\begin{define}\label{def:HW}
An Execution Hardware Architecture $H^A = \langle E, L \rangle$ comprises execution nodes $E$ and communication links $L = E \times E$ between these nodes. The set of execution nodes $E = E^C \cup E^A$ is composed by a set of central execution nodes $E^C = \lbrace e_1 ,..., e_k \rbrace$ and a set of peripheral Smart-Aggregate nodes $E^A = \lbrace e_{k+1} ,..., e_l \rbrace$ with attached physical Sensors and Actuators. The set $E^C$ is also called the \emph{Central Platform Computer} (CPC).
\end{define}

\begin{define}\label{def:DeploymentConfiguration}
The Configuration $\Phi = \langle \delta_P(SC),$ $\delta_A(SC),$ $\delta(SC) \rangle$ defines how ASWC-Clusters $SC$ are deployed to execution nodes $E$, either passively ($\delta_P$) or actively ($\delta_A$). For $sc \in SC$, we define 
$\delta_P(sc)\xrightarrow[]{} \lbrace e_i \in E \vbar sc$ is in memory of $e_i$, but not executed on $e_i \rbrace$,
$\delta_A(sc)\xrightarrow[]{} \lbrace e_i \in E \vbar sc$ is in memory of $e_i$ and executed on $e_i \rbrace$ and
$\delta(sc) = \delta_A(sc) \cup \delta_P(sc)$.
\end{define}

Our deployment approach can either be applied to ASWCs or to ASWC-Clusters. The motivation to think in Clusters and not in single ASWCs is that the definition of Clusters reduces the complexity with regard to the amount of combinations to be considered for deployment and master-slave switchovers. Furthermore, the ASWCs within a Cluster have a kind of stronger binding to each other. Thus, we aim on a deployment of ASWCs which are bound to one cluster within the same ECU. An example for a binding quality is data-transport delay.

ASWCs might contain invisible sub-components and internal communication channels. We don't model external communication channels between ASWCs in this paper for simplicity.

\subsection{Fixed Properties of the Deployment Model} 
\label{subsec:fixedParams}

Each ASWC $s_i \in S$ is defined by several properties.
Property $wcet(S) \toSetN^+$ defines the \emph{Worst-Case Execution Time}. 
Property $asil(S) \toSet[0..4]$ defines the \emph{Automotive Safety Integrity Level} (ASIL) of an ASWC [0: Quality-Management (QM), 1: ASIL-A, 2: ASIL-B, 3: ASIL-C, 4: ASIL-D].
Property $failOp(S) \toSetN_0$ defines the fail-operational level [0: non fail-operational, $n$: $s_i$ has to be provided after $n$ isolations].
The minimum of the fault-tolerance times of an ASWC for its different safety goals is defined by $minFTT(S) \toSetN^+$.

As defined in section \ref{subsec:redundancyConcepts},  the vehicle property $frt(\Veh) \toSetN^+$ defines the \emph{fault-recovery time} of the vehicle $\Veh$. 
The $frt$ has influence on if the slaves are deployed as hot or as cold-slaves, depending on their $minFTT$.

For execution nodes $e \in E$, the following properties are defined.
The property $totalTimeBudget(E) \toSetN^+$ defines the budget of time that is provided in each cycle to execute the ASWCs. We assume here that ASWCs are executed in every cycle.
The property $powerSupply(E) \toSet[0,1]$ defines the power supply of the execution node [0: Blue, 1: Red].
Finally, the property $isolated(E) \toSetBool$ defines if the execution node $e_i \in E$ is isolated in the current solution instance.
We do not model the amounts of required and provided volatile and non-volatile memory here for simplicity. These are handled in a similar manner than the WCET and the time-budget.

\subsection{Solution Properties of the Model} 
\label{subsec:solutionVariables}

In this section we describe the model-properties that represent the solution of the deployment problem.

The properties of ASWC-Clusters $sc \in SC$ depend on the mapped ASWCs.
Properties $asil(SC) \toSet[0..4]$ and $failOp(SC) \toSetN_0$ define the ASIL and the fail-operational level of a cluster.
It's ensured by constraints that $\forall s_i \in \alpha(sc): asil(sc) = asil(s_i)$ and $failOp(sc) = failOp(s_i)$.
Property $minFTT(SC) \toSetN^+$  is the minimum of all the $minFTT(s_i)$ for $s_i \in \alpha(sc)$. 
The property $sumWcets(SC)$ is defined to be equal to $\sum_{s_i \in \alpha(sc)} wcet(s_i)$.

For execution nodes $e \in E$, $usedTimeBudget(E) \toSetN_0$ is defined to be equal to
$\sum_{sc_j \in SC \vbar e \in \delta_A(sc_j)} sumWcets(sc_j)$, which is the sum of the $wcet(s)$ of those ASWCs that are active in the schedule on node $e$. 
A constraint ensures that $\forall e \in E : usedTimeBudget(e) \leq totalTimeBudget(e)$.

Notice that the decision if an ASWC-Cluster instance becomes a master or a hot-standby slave is done at runtime by a Platform-Management component of the RTE of the proposed vehicle platform. 
This is, because there are also other reasons beside node-isolations that may lead to the deactivation of a master. 
Hence, the calculated master/slave deployments as shown in this paper are not used as predefined runtime-configuration, but at design-time to statically analyze the fail-operational runtime-behavior. 
It can be analyzed under which circumstances it is possible at runtime to keep a master resp. a slave alive in the presence of faults that lead to node-isolations.

\subsection{Deployment Constraints and Problem Solving} 
\label{subsec:constraints}

To define the set of valid deployments, we setup an arithmetic model of the system properties and the deployment constraints.
We implemented the deployment calculation and analysis by defining the model with arithmetic calculations and simple functions like \emph{Sums}, \emph{Implications} and \emph{if/then/else} relations. We do not list the detailed constraints in this paper for space reasons. The model can be solved for instance by an SMT-Solver like Z3 \cite{demoura2008z3}.

\subsection{Example of an Initial Deployment}
\label{sec:example_initial}

In this section we show an application of our approach on a simplified example from the automotive domain. Consider the following functional features and ASWCs:


~\\
\begin{tabular}{|l@{\hspace{4pt}}|l@{\hspace{4pt}}|l|} 
\hline
										&																& \textbf{asil($s_i$)} / \\
\textbf{Feature $f_i$}		& 	\textbf{ASWCs} $s_i$ of $\chi(f_i)$ 		& \textbf{failOp($s_i$)} / \\
										&																& \textbf{wcet($s_i$)} in ms \\
\hline
$f_1:$ Infotainment 		& $s_1:$ Infotainment      						& QM / 0 / 2 \\ 
\hline 
$f_2:$ Energy- 				& $s_2:$ RemainingDrive-     					& A  / 0  / 0.7 \\ 
\;\;\;\;\;\;\;Management	& \;\;\;\;\;\;\;RangeEstimation 				& \\ 
										& $s_3:$ EnergyEfficiency- 						& A /  0 / 0.3 \\ 
										& \;\;\;\;\;\;\;Assistant 							& \\ 

\hline 
$f_3:$ ADAS-A  				& $s_4:$ AdasSwc1 									& C / 0  / 1.7 \\ 
               							& $s_5:$ AdasSwc2 									& D / 1  / 1 \\ 
\hline 
$f_4:$ ADAS-B 				& $s_5:$ AdasSwc2									& D / 1 / 1 \\ 
\hline 
$f_5:$ Manual 					& $s_6:$ ManualAccelerate 						& D / 3 / 1 \\ 
\;\;\;\;\;\;\;Driving			& $s_7:$ ManuelBrake 								& D / 3 / 1 \\ 
										& $s_8:$ ManualSteer 								& D / 3 / 0.5 \\ 
\hline
\end{tabular}


The features $f_3$ and $f_4$ are placeholders for some Advanced Driver Assistance Systems (ADAS), like an ACC or automatic parking. 
Notice that feature $f_3$ is realized by two ASWCs with different properties and ASWC $s_5$ contributes to realize two features $f_3$ and $f_4$. 
Feature $f_3$ is non fail-operational, as not all $s_i \in \chi(f_3)$ are fail-operational, but $f_4$ is fail-operational.

In this example, five ASWC-Clusters  $\lbrace sc_1, ..., sc_5 \rbrace$ are established. 
Due to the constellation of the properties $asil(s_i)$ and $failOp(s_i)$, the ASWC-Clusters are: 
$\alpha(sc_1) = \lbrace s_1 \rbrace$, $
\alpha(sc_2) = \lbrace s_2, s_3 \rbrace$, $
\alpha(sc_3) = \lbrace s_4 \rbrace$, $
\alpha(sc_4) = \lbrace s_5 \rbrace$ and $
\alpha(sc_5) = \lbrace s_6, s_7, s_8 \rbrace$.
Notice that ASWC $s_5$ is only in one cluster, although it contributes to two features.

\begin{figure}[h!]
\centering
\includegraphics[width = 0.48\textwidth]{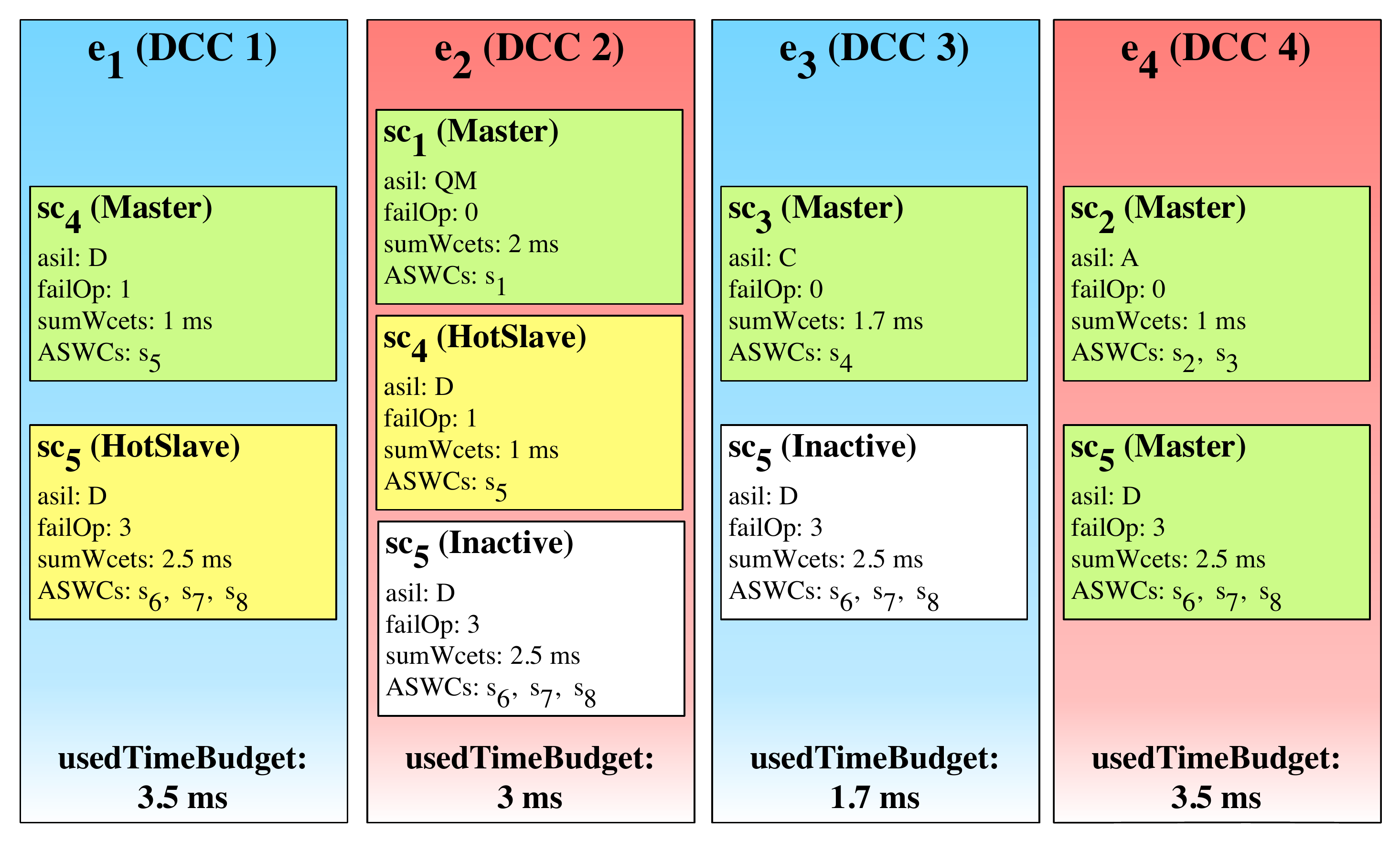}
\caption{Initial deployment for the example}
\label{fig:Example_Initial}
\end{figure}

Considering a CPC with 4 execution nodes (DCCs) as shown in Fig. \ref{fig:RACE_Platform}, a valid initial deployment for the example is shown in Fig. \ref{fig:Example_Initial}. 
The colors (red/blue) of the execution nodes denote their attached power-supply.
We assume here that $minFTT(s_i) < frt(\Veh)$ for all fail-operational ASWCs. 
Hence, hot-standby slaves are required.
As provided execution time of the execution nodes per cycle, we assume $totalTimeBudget(e_i) = 4 ms$. 
As visible in Fig. \ref{fig:Example_Initial} at the values of $usedTimeBudget(e_i)$, no time-budget of any execution node is exceeded.

The Z3 SMT-Solver \cite{demoura2008z3} calculated the initial deployment in 125ms on a 2 GHz Core i7. 
This is the time for the check()-operation, not the time for setting up the model and constraints.

\subsection{Reconfigurations after Isolations}
\label{subsec:ReconfigurationsAfterIsolations}

Let $E^C_f \subset E^C$ be the set of isolated execution nodes. For all $e_i \in E^C_f$, we set $isolated(e_i)=1$. It is ensured by constraints that no ASWC-Cluster is activated anymore on one of the isolated execution nodes.

\begin{define}\label{def:PAG}
A Platform-Availability-Graph (PAG) is a directed acyclic graph $G=(V,E)$. 
Each vertex $V$ represents a set of alive central execution nodes $E^C_a = E^C \setminus E^C_f$. 
The edges $E$ describe a transition between two vertexes, meaning that some $e_i \in E^C$ move from $E^C_a$ to $E^C_f$.
A transition happens due to an isolation or if a power-supply disappears.
\end{define}

\begin{figure}[h!]
\centering
\includegraphics[width=2in]{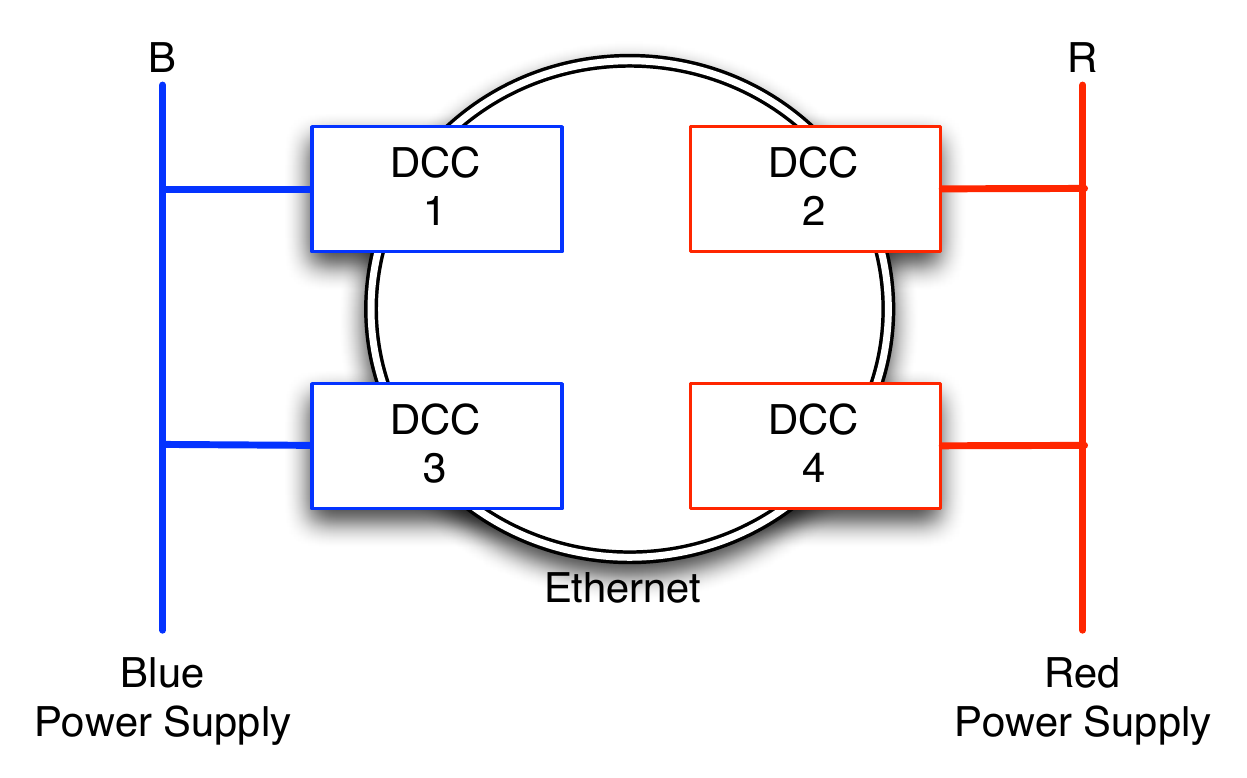}
\caption{An example Central Platform Computer (CPC) with 4 DCCs}
\label{fig:RACE_PAG_CPC}
\end{figure}

Fig. \ref{fig:RACE_PAG_CPC} shows an example CPC containing 4 central execution nodes (DCCs) and the two power-supplies (red and blue).

\begin{figure}[h!]
\centering
\includegraphics[width=2.5in]{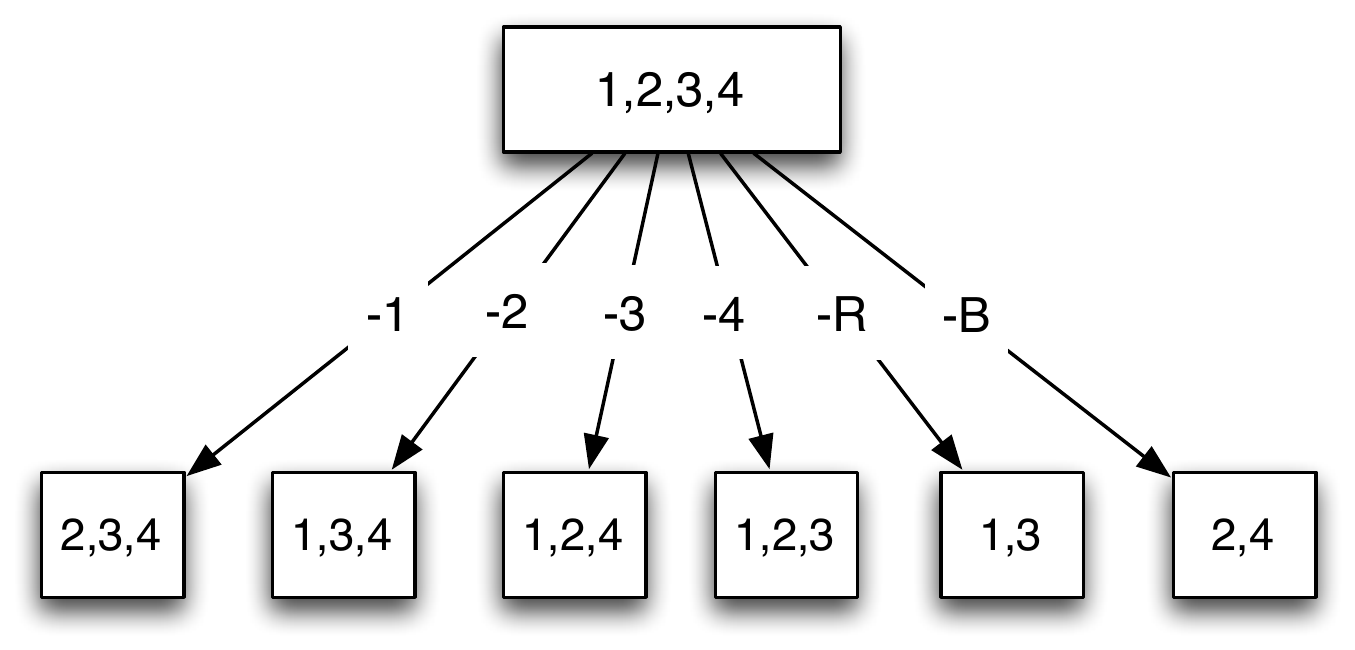}
\caption{Example PAG considering only one fault}
\label{fig:RACE_PAG_4DCCs_1Failure}
\end{figure}

When considering only one fault, the PAG looks like shown in Fig. \ref{fig:RACE_PAG_4DCCs_1Failure}. 
The vertexes are labeled with the Ids $i$ of the alive nodes $e_i \in E^C_a$.
The edges are labeled with the Id $i$ of that $e_i \in E^C_f$ which has recently been isolated resp. with the power-supply ($R, B$) that has recently been broken down.

The validity of deployments after a transition in the PAG is ensured by arithmetic constraints. For instance, these constraints ensure that the mapping of ASWCs to clusters is not changed during a PAG-transition. Furthermore, it is ensured that the previous allocations or activations of ASWC-Clusters to execution nodes are not changed unnecessarily during a PAG-transition.
To do this, some solution properties of the former deployment are used as fixed properties for the follow-up deployment, when a PAG-transition is calculated. 

To cover deactivation scenarios that might be required after isolations of central execution nodes, each $sc \in SC$ has additionally the following properties:

\begin{itemize}
\item 
$hotStandbySlaveReq(SC) \toSetBool$:\quad
indicates if a hot-standby \emph{slave} is required. The valuation is derived by considering $minFTT(sc)$  and $frt(\Veh)$
\item 
$hotStandbySlavePresent(SC) \toSetBool$:\quad
indicates if a required hot-standby \emph{slave} can be established
\item 
$masterPresent(SC) \toSetBool$:\quad
indicates if the \emph{master} can be established or not
\end{itemize}

In order to decide about the deactivation order for the ASWC-Clusters, each ASWC-Cluster has assigned the properties $prioPointsMaster(SC) \toSetN^+$ and $prioPointsHotSlave(SC) \toSetN^+$ storing priorities of actively deployed instances of the cluster. 

We calculate the sum of the priorities of all active instances of ASWC-Clusters and use these priorities and their sum to construct an order in which the instances of the clusters should be deactivated in case system resources become insufficient. 
We do this by maximizing the sum of the priorities.
The clusters with the lowest priority get deactivated first.  
We derive the priorities depending on $asil(SC)$ and $failOp(SC)$.
We set 
$prioPointsMaster(sc) = asil(sc) + failOp(sc) + 2$ and 
$prioPointsHotSlave(sc) = asil(sc) + failOp(sc) + 1$.
Hence, clusters with lowest ASIL will get deactivated first.
However, the priorities could also be set differently. 

\subsection{Example of a Deployment after an Isolation}
\label{sec:example_isolation}

\begin{figure}[h]
\centering
\includegraphics[width = 0.48\textwidth]{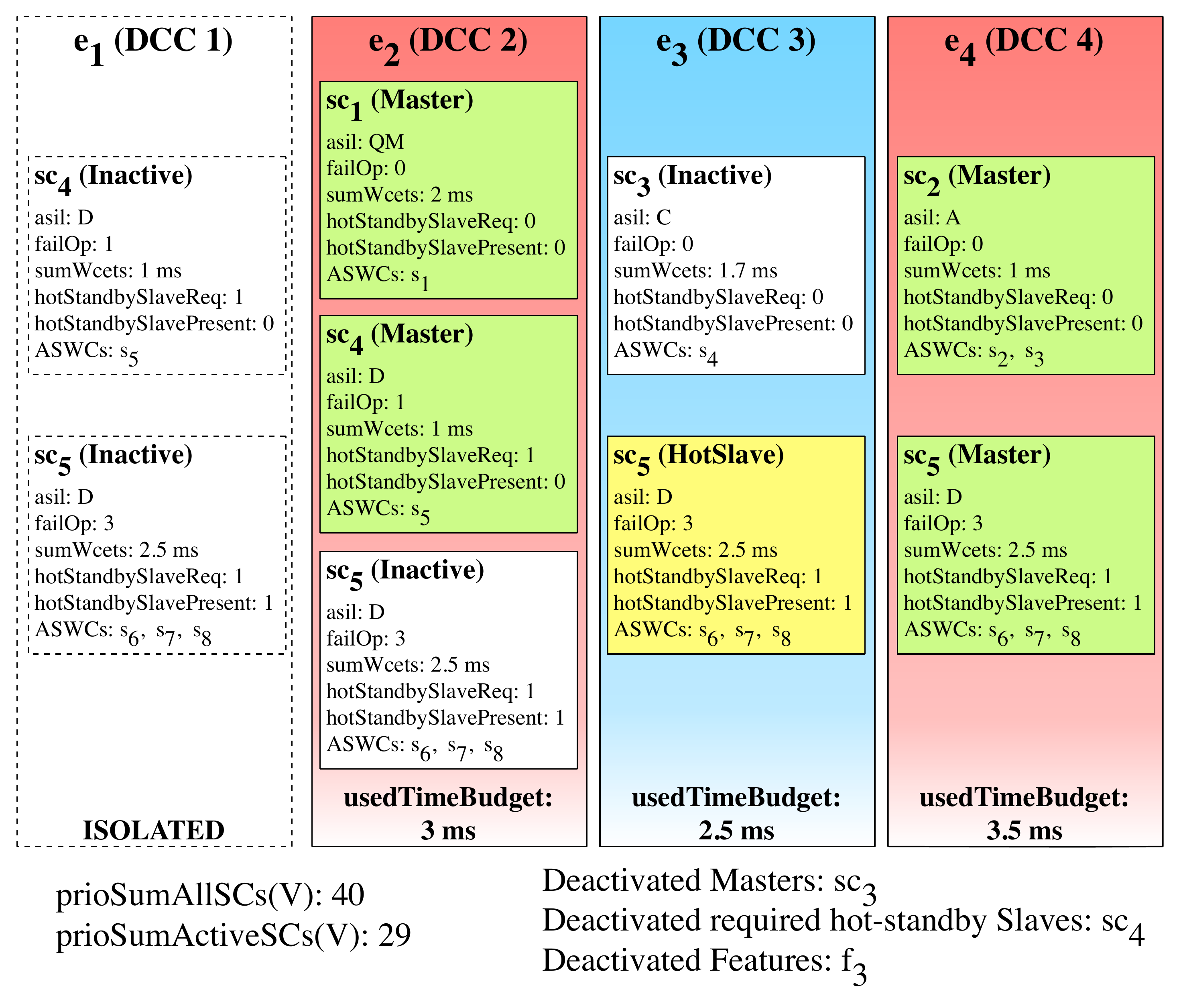}
\caption{Follow-up deployment after DCC1 has been isolated}
\label{fig:Example_IsolatedDCC1}
\end{figure}

Fig. \ref{fig:Example_IsolatedDCC1} shows the follow-up deployment for the case that DCC1 becomes isolated in the initial deployment (cf. Fig. \ref{fig:Example_Initial}).

While in the initial deployment all clusters can be deployed as required, after the isolation of $e_1$ (= DCC 1) the master of cluster $sc_4$ gets lost and its slave on $e_2$ becomes the new master. 
As $failOp(sc_4)=1$, no new slave is created because it's not required that $sc_4$ is still present after the next isolation.
Furthermore, the slave of cluster $sc_5$ get lost.
As $failOp(sc_5)=3$, an inactive instance of $sc_5$ must be activated to serve as new slave to prepare for the next isolation.
The new slave of $sc_5$ can only be activated on $e_3$, not on $e_2$, because it is not allowed that master and slave depend on the same power-supply. 
However, to be able to execute $sc_5$ on $e_3$, $sc_3$ has to be deactivated as the sum of the WCETs of $sc_3$ and $sc_5$ would exceed the time-budget of $e_3$. 
The deactivation of $sc_3$ forces the deactivation of feature $f_3$, as $\alpha(sc_3) = \lbrace s_4 \rbrace \subseteq \chi(f_3)$.

The sum of priority points in the initial solution was 40. 
The loss of the master of $sc_3$ and the slave of $sc_4$ forces a loss of 11 priority points ($prioPointsMaster(sc_3) = 5$, $prioPointsHotSlave(sc_4) = 6$). 
Hence, without DCC1 only 29 priority points can be provided by the system (cf. Fig. \ref{fig:Example_IsolatedDCC1}).

When this procedure is continued by isolating more DCCs in arbitrary order, the cluster $sc_5$ always has a master instance, even if only one DCC is left. This is important as $failOp(sc_5) = 3$.

To calculate the follow-up deployment, 1.1s were spent for checking 11 models that are unsatisfiable for priority point sums 40 to 30, plus 90ms for checking the valid solution.

The designer can analyze the system's fail-operational behavior by considering the set of deactivated features for each situation. This allows to formally analyze if all desired system and feature properties can be fulfilled, without executing the system.
The initial deployment can also be changed manually in order to analyze the systems feature availability depending on different initial deployments.

\section{Related Work} 
\label{sec:rw}

In this section, we discuss related work of deployment approaches with focus on safety and fail-operationality.


In \cite{shelton2003framework}, the authors show an approach to analyze graceful degradation.
They use a utility function to measure the set of active features.
This can be seen as quite similar to our sums of priorities.
To reduce complexity, they group components by defining subsystems based on the interfaces of components.
We group components by their dependability requirements. 
This allows separation of mixed-critical components.
The main differences are that they consider a fail-silent fault-model, while we consider fail-operational behavior of features.
Furthermore, we focus more explicitly on deployment constraints that ensure fail-operational behavior.
Another difference is that we consider the explicit deactivation of components to be able to keep alive other components that are required to behave fail-operational.
They consider a fixed hardware configuration, while we consider a HW-Architecture whose provided resources decrease after random hardware failures due to execution node isolations.


In \cite{pinello2008fault}, the authors introduce a design methodology for safety-critical systems, called SCRAPE (Safety-Critical Real-Time APlications Exploration). In addition the \emph{fault-tolerant data flow} (FTDF) model of computation is introduced. The SCRAPE design flow has 6 main steps. Our work is mainly related to steps 4 and 5, namely the specification of \emph{Fault Behavior and Mapping Constraints} as well as the calculation of a \emph{Fault-Tolerant Embedded Software Deployment}. However, with SCRAPE fail-silent execution platforms are addressed, while we focus on ensuring fail-operational features. Additionally, we address the analysis about which feature-set can be provided after certain random hardware failures.


In \cite{boone2008automated}, fault-tolerant deployments with focus on the trade-off between Performance and Reliability are optimized using a MILP-Solver.
However, the approach does not consider mixed criticalities explicitly, and also at most 1 replication is supported due to the single node failure model. The analysis of deployments after hardware-faults is also not considered.


In \cite{armbruster2009fahrzeugubergreifende}, a dynamically reconfigurable vehicle control platform supporting fault-tolerance is described. The reactivation- and reallocation algorithms are executed during runtime as one of the platform's core-algorithms. However, only two clusters of application components are supported. It is ensured that at any time the most important cluster is executed with a fail-operational behavior and that the last non-faulty execution node executes the most important cluster. Anyhow, the presented work is limited to only two clusters whereas one does not have any fail-operational requirement.


\section{Conclusion and Future Work} 
\label{sec:conclusion}

In this paper, we have shown a formal approach to calculate deployments of mixed-critical functional features in a new HW/SW architecture for vehicles. 
The work allows to analyze the fail-operational behavior of features in the presence of random hardware failures.
It can be analyzed which features can be uphold depending on the available set of execution nodes.
We defined a formal system model and ensured deployment constraints by setting up an arithmetic input model for a SMT-Solver, that calculates valid deployments.

As future work, we are going to include communication channels between ASWCs into the model.
In this context, we are going to select optimal channels out from a set of channel-candidates and optimize the deployment respective to minimize the required network bandwidth. 
An end-2-end timing-analysis with focus on the deployment options is also planned.

Furthermore, we want to treat the integration of new software components into existing deployments during the use case of extensions of the vehicle by new functional features in a plug-and-play manner. 
Finally, we want to evaluate the scalability of our approach based on the system layout of a concept car that we construct.


\section*{Acknowledgment}

This work is partially funded by the German Federal Ministry for Economic Affairs and Energy (BMWi) under grant no. 01ME12009 through the project RACE (Robust and Reliant Automotive Computing Environment for Future eCars) (http://www.projekt-race.de).



%

\bibliographystyle{IEEEtran}
\bibliography{literature}

\begin{thebibliography}{10}
\providecommand{\url}[1]{#1}
\csname url@samestyle\endcsname
\providecommand{\newblock}{\relax}
\providecommand{\bibinfo}[2]{#2}
\providecommand{\BIBentrySTDinterwordspacing}{\spaceskip=0pt\relax}
\providecommand{\BIBentryALTinterwordstretchfactor}{4}
\providecommand{\BIBentryALTinterwordspacing}{\spaceskip=\fontdimen2\font plus
\BIBentryALTinterwordstretchfactor\fontdimen3\font minus
  \fontdimen4\font\relax}
\providecommand{\BIBforeignlanguage}[2]{{%
\expandafter\ifx\csname l@#1\endcsname\relax
\typeout{** WARNING: IEEEtran.bst: No hyphenation pattern has been}%
\typeout{** loaded for the language `#1'. Using the pattern for}%
\typeout{** the default language instead.}%
\else
\language=\csname l@#1\endcsname
\fi
#2}}
\providecommand{\BIBdecl}{\relax}
\BIBdecl

\bibitem{avizienis2004basic}
A.~Avizienis, J.~Laprie, B.~Randell, and C.~Landwehr, ``Basic concepts and
  taxonomy of dependable and secure computing,'' \emph{IEEE Trans. on
  Dependable and Secure Computing}, vol.~1, no.~1, pp. 11--33, 2004.

\bibitem{chakraborty2012embedded}
S.~Chakraborty, M.~Lukasiewycz, C.~Buckl, S.~Fahmy, N.~Chang, S.~Park, Y.~Kim,
  P.~Leteinturier, and H.~Adlkofer, ``Embedded systems and software challenges
  in electric vehicles,'' in \emph{Proceedings of the Conference on Design,
  Automation and Test in Europe}.\hskip 1em plus 0.5em minus 0.4em\relax EDA
  Consortium, 2012, pp. 424--429.

\bibitem{sommer2013race}
S.~Sommer, A.~Camek, K.~Becker, C.~Buckl, A.~Knoll, A.~Zirkler, L.~Fiege,
  M.~Armbruster, and G.~Spiegelberg, ``Race: A centralized platform computer
  based architecture for automotive applications,'' in \emph{IEEE Vehicular
  Electronics Conference / Int. Electric Vehicle Conference (VEC-IEVC)}, 2013.

\bibitem{henzinger2001giotto}
T.~Henzinger, B.~Horowitz, and C.~Kirsch, ``Giotto: A time-triggered language
  for embedded programming,'' in \emph{Embedded Software}.\hskip 1em plus 0.5em
  minus 0.4em\relax Springer, 2001, pp. 166--184.

\bibitem{iso26262_1}
{International Organization for Standardization}, ``{ISO/DIS 26262-1 - Road
  vehicles - Functional safety, Part 1 Glossary},'' Technical Committee 22
  (ISO/TC 22), Geneva, CH, Tech. Rep., Nov. 2011.

\bibitem{kopetz2003fault}
H.~Kopetz, ``Fault containment and error detection in the time-triggered
  architecture,'' in \emph{Int. Symposium on Autonomous Decentralized Systems
  (ISADS)}.\hskip 1em plus 0.5em minus 0.4em\relax IEEE, 2003, pp. 139--146.

\bibitem{blanke2001concepts}
M.~Blanke, M.~Staroswiecki, and N.~E. Wu, ``Concepts and methods in
  fault-tolerant control,'' in \emph{Proceedings of the 2001 American Control
  Conference}, vol.~4.\hskip 1em plus 0.5em minus 0.4em\relax IEEE, 2001, pp.
  2606--2620.

\bibitem{demoura2008z3}
L.~De~Moura and N.~Bj{\o}rner, ``Z3: An efficient smt solver,'' \emph{Tools and
  Algorithms for the Construction and Analysis of Systems}, pp. 337--340, 2008.

\bibitem{shelton2003framework}
C.~Shelton, P.~Koopman, and W.~Nace, ``A framework for scalable analysis and
  design of system-wide graceful degradation in distributed embedded systems,''
  in \emph{Int. Workshop on Object-Oriented Real-Time Dependable Systems
  (WORDS)}.\hskip 1em plus 0.5em minus 0.4em\relax IEEE, 2003, pp. 156--163.

\bibitem{pinello2008fault}
C.~Pinello, L.~P. Carloni, and A.~L. Sangiovanni-Vincentelli, ``Fault-tolerant
  distributed deployment of embedded control software,'' \emph{IEEE
  Transactions on Computer-Aided Design of Integrated Circuits and Systems},
  vol.~27, no.~5, pp. 906--919, 2008.

\bibitem{boone2008automated}
B.~Boone, F.~De~Turck, and B.~Dhoedt, ``Automated deployment of distributed
  software components with fault tolerance guarantees,'' in \emph{6th Int.
  Conf. on Software Engineering Research, Management and Applications
  (SERA)}.\hskip 1em plus 0.5em minus 0.4em\relax IEEE, 2008, pp. 21--27.

\bibitem{armbruster2009fahrzeugubergreifende}
M.~Armbruster, ``Eine fahrzeug{\"u}bergreifende x-by-wire plattform zur
  ausf{\"u}hrung umfassender fahr-und assistenzfunktionen,'' Ph.D.
  dissertation, Institute for Avionics Systems (Institut f{\"u}r
  Luftfahrtsysteme, ILS), University of Stuttgart, 2009.

\end{thebibliography}



\end{document}